# Neutrino spin oscillations in gravitational fields

**S. A. Alavi [1]; S. F. Hosseini [1,2]**

[1] *Department of Physics, Hakim Sabzevari University, P. O. Box 397, Sabzevar, Iran.*
[2] *Education management organization, Quds city, P. O. Box 1187613911, Tehran, Iran.*

*s.alavi@hsu.ac.ir* ; *alialavi@fastmail.us* ; *sfho3ini@yahoo.com*

*We study neutrino spin oscillations in black hole backgrounds. In the case of a charged black hole, the maximum frequency of oscillations is a monotonically increasing function of the charge. For a rotating black hole, the maximum frequency decreases with increasing the angular momentum. In both cases, the frequency of spin oscillations decreases as the distance from the black hole grows. As a phenomenological application of our results, we study simple bipolar neutrino system which is an interesting example of collective neutrino oscillations. We show that the precession frequency of the flavor pendulum as a function of the neutrino number density will be higher for a charged/non-rotating black hole compared with a neutral/rotating black hole respectively.*

**PACS No:** 14.60.Pq, 14.60.St, 04.25.Nx, 04.25.-g

**Key words:** neutrino spin oscillations, gravitational fields, Reissner-Nordstrom (RN) and Kerr metrics.

## Introduction

Neutrino physics is an active area of research with important implications for particle physics, cosmology and astrophysics. Cosmological implications of neutrinos include lepto/ baryogenesis and possible connections to the dark sector of the universe [1]. The phenomenon of neutrino oscillations can explain solar and atmospheric neutrino problems. It also provides the first experimental evidence for physics beyond the standard model since it requires nonzero mass for neutrinos.

The effects of gravitational fields on the neutrino oscillations have been studied in the literature. In [2], neutrino flavor oscillations in the Kerr-Newman space-time has been studied. In [3], the authors investigated the effect of a Quantum Gravity-induced minimal length on neutrino oscillations. Neutrino optics in gravitational fields has been studied in [4]. Interaction of neutrinos with an external field provides one of the factors required for a transition between helicity states. In [5], neutrino spin oscillations has been studied in Schwarzschild background which describes the gravitational field of an uncharged and non-rotating black hole.

In this paper, we study neutrino spin oscillations in Reissner-Nordstrom (RN) and Kerr backgrounds which describe the gravitational fields of a charged non-rotating and a rotating black hole respectively.

## Neutrino spin oscillations in the Reissner-Nordstrom (RN) metric

RN metric describes the gravitational field of a charged non-rotating black hole.

$$d\tau^2 = A^2 dt^2 - A^{-2} dr^2 - r^2 d\theta^2 - r^2 \sin^2\theta d\varphi^2, \tag{1}$$

$$A = \sqrt{1 - \frac{2M}{r} + \frac{Q^2}{r^2}}. \tag{2}$$

Here $Q$ and M are the charge and the mass of the black hole respectively and we have used the natural units $\hbar = c = 1$. The components of vierbein four velocity are [5]:

$$u^a = (\gamma A, U_r A^{-1}, U_\theta r, U_\varphi r \sin\theta), \tag{3}$$

where

$$U^\mu = (U^0, U_r, U_\theta, U_\varphi), \ U^0 = \frac{dt}{d\tau} = \gamma \ , U_r = \frac{dr}{d\tau}, U_\theta = \frac{d\theta}{d\tau}, U_\varphi = \frac{d\varphi}{d\tau}, \tag{4}$$

is the four velocity of a particle in its geodesic path, which is related to $u^a$ through $u^a = e^a_\mu U^\mu$. The four velocity of a particle in the relevant metric $U^\mu$ is related to the world velocity of the particle through $\vec{U} = \gamma \vec{V}$ where $\gamma = \frac{dt}{d\tau}$ and $\tau$ is the proper time. The nonzero vierbein vectors are :

$$e^0_t = (1 - \frac{2M}{r} + \frac{Q^2}{r^2})^{\frac{1}{2}}, \ e^1_r = (1 - \frac{2M}{r} + \frac{Q^2}{r^2})^{-\frac{1}{2}}, \ e^2_\theta = r \ , \ e^3_\varphi = r \sin\theta. \tag{5}$$

To study the spin evolution of a particle in a gravitational field, we calculate $G_{ab} = (\vec{E}, \vec{B})$ which is the analogue of the electromagnetic field tensor. It is defined as follows:

$$G_{ab} = e_{a\mu;\nu} e^\mu_b U^\nu, \tag{6}$$

where $e_{a\mu;\nu}$ are the covariant derivatives of vierbein vectors given by :

$$e_{a\mu;\nu} = \frac{\partial e_{a\mu}}{\partial x^\nu} - \Gamma^\lambda_{\mu\nu} e_{a\lambda}. \tag{7}$$

After using Eqs.(5) and (7) to calculate the covariant derivatives of vierbein vectors, we have:

$$e_{0r;t} = -\frac{1}{2}\left(\frac{2M}{r^2} - \frac{2Q^2}{r^3}\right) A^{-1},$$

$$e_{1t;t} = \frac{1}{2}\left(\frac{2M}{r^2} - \frac{2Q^2}{r^3}\right) A, \ e_{1\theta;\theta} = \left(-r + 2M - \frac{Q^2}{r}\right) A^{-1}, \ e_{1\varphi;\varphi} = \sin^2\theta A^{-1}\left(-r + 2M - \frac{Q^2}{r}\right),$$

$$e_{2r;\theta} = 1 , e_{2\varphi;\varphi} = -r \sin\theta \cos\theta,$$

$$e_{3r;\varphi} = \sin\theta , e_{3\theta;\varphi} = r \cos\theta. \tag{8}$$

We recall that an antisymmetric tensor in four dimensions can be stated in terms of two three dimensional vectors (such as electric and magnetic fields):

$$G_{ab} = (\vec{E}, \vec{B}) , G_{0i} = E_i , G_{ij} = -\epsilon_{ijk} B_k. \tag{9}$$

It is worth to mention that the electric and magnetic fields in (9), are not (real)electric and magnetic fields but they are only their analogues.

Using Eqs.(4), (5), (6), (8), and (9), we have the following forms for the electric and magnetic fields:

$$\vec{E} = \left(-\frac{\gamma}{2}\left(\frac{2M}{r^2} - \frac{2Q^2}{r^3}\right), 0, 0\right) , \vec{B} = (\gamma v_\varphi \cos\theta, -\gamma v_\varphi A \sin\theta, \gamma A v_\theta). \tag{10}$$

Since the gravitational field around a charged non-rotating black hole is symmetric, we can consider neutrino motion in the equatorial plane $\theta = \frac{\pi}{2}$ (hence $U_\theta = 0$). The four-velocity in the vierbein frame is not constant in general case but it can be shown that for circular orbits $\frac{du^a}{d\tau} = 0$, which means that the velocity four vector of neutrino is constant with respect to the vierbein frame. We assume that the motion is in a circular orbit with constant radius $r$ ($U_r = \frac{dr}{d\tau} = 0$). It is important to note that not all the orbits with arbitrary radius are stable. Here we concentrate on the stable orbits. The geodesic equation of a particle in a gravitational field is described by [6]:

$$\frac{d^2 x^\mu}{dq^2} + \Gamma^\mu_{\sigma\nu} \frac{dx^\sigma}{dq} \frac{dx^\nu}{dq} = 0, \tag{11}$$

where variable $q$ parameterizes the particle's world line. From Eqs. (1), (2), and (11), we can calculate the values of $v_\varphi$ and $\gamma^{-1}$:

$$\gamma^{-1} = \sqrt{1 - \frac{3M}{r} + \frac{2Q^2}{r^2}}, \tag{12}$$

$$v_\varphi = \sqrt{\frac{M}{r^3} - \frac{Q^2}{r^4}}. \tag{13}$$

Neutrino spin precession is given by the expression $\vec{\Omega} = \frac{\vec{G}}{\gamma}$, where vector $\vec{G}$ is defined as follows [7]:

$$\vec{G} = \frac{1}{2}\left[\vec{B} + \frac{1}{1+u^0}[\vec{E} \times \vec{u}]\right]. \tag{14}$$

After substituting (3), (10), (12), (13), (14) and $\vec{U} = \gamma\vec{V}$ in $\vec{\Omega} = \frac{\vec{G}}{\gamma}$, we obtain the only nonzero component of frequency :

$$\Omega_2 = -\frac{1}{4M}\sqrt{1 - \frac{3M}{r} + \frac{2Q^2}{r^2}}\sqrt{\frac{4M^3}{r^3} - \frac{4M^2Q^2}{r^4}} = -\frac{v_\varphi}{2}\gamma^{-1}. \tag{15}$$

In Fig.(1) we show $2|\Omega_2|M$ versus $\frac{r}{2M}$ for different values of $\alpha = \frac{Q}{M}$. For $\alpha = 0$ our results reduce to that in the case of the Schwarzschild metric [5].

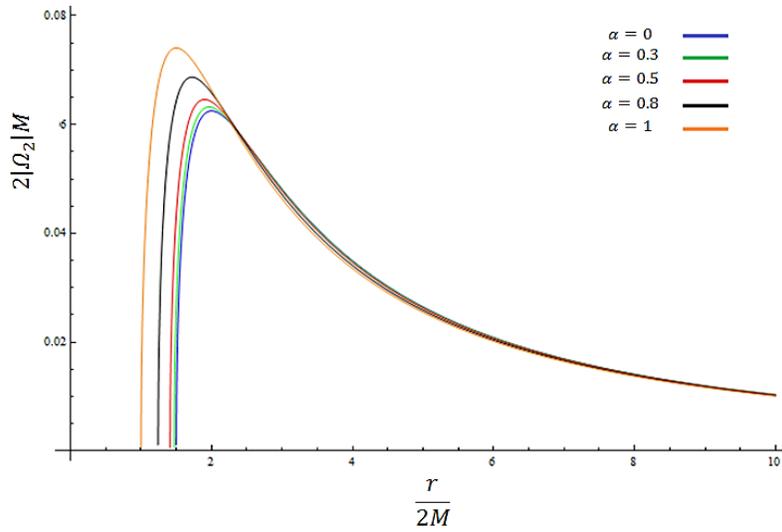

Figure 1: Neutrino spin oscillation frequency versus the radius of the neutrino orbit for $0 \leq \alpha \leq 1$.

We stop at $\alpha = 1$ since the RN metric with $Q > M$ does not describe a black hole due to the existence of a naked singularity at $r=0$.

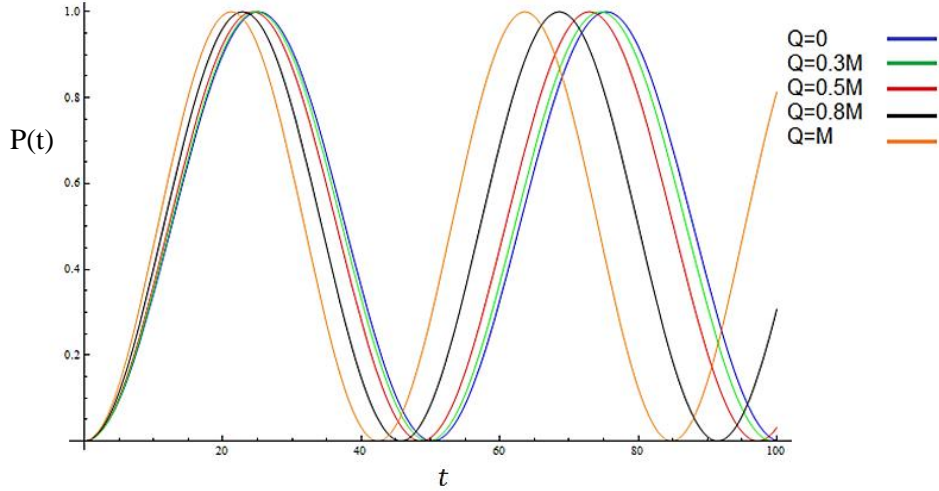

Figure 2: The transition probability versus time for different values of $Q$.

It is seen that the maximum frequency of oscillations increases with $\alpha$. It is also seen that $\Omega_2$ has its largest value for $Q = M$. $\alpha = 0$ curve is in agreement with the result obtained for the Schwarzschild metric [5]. We also see that curves for different values of $\alpha$ coincide at large $r$. This indicates that the effect of the black hole charge diminishes at long distances. In Fig.(2) we show the neutrino transition probability $P(t) = sin^2(\Omega_2 t)$.

## Neutrino spin oscillations in the Kerr metric

Kerr metric [8] describes the geometry of space-time in the vicinity of a rotating balck hole. The Kerr metric is usually written in Boyer-Lindquist coordinates, which is not diagonal. One can transform the Kerr metric in to the standard form (diagonal form) with an appropriate change of variable [9].

$$ds^2 = -Cdt^2 + Fdr^2 + Rd\varphi^2 + Hd\theta^2,$$

(16)

where :

$$H = \rho^2, \; C = \frac{\Delta}{H}, \; R = \frac{S}{H}, \; F = \frac{H}{\Delta},$$

(17)

and :

$$a = \frac{J}{M}, \; \rho^2 = r^2 + a^2\cos^2\theta, \; \Delta = r^2 + a^2 - 2Mr, \; S = (r^2 + a^2)^2 \sin^2\theta.$$

(18)

The vierbein vectors and their inverses are as follows:

$$e_\mu^0 = (\sqrt{C}, 0, 0, 0), e_\mu^1 = (0, \sqrt{F}, 0, 0), e_\mu^2 = (0, 0, \sqrt{H}, 0), e_\mu^3 = (0, 0, 0, \sqrt{R}),$$

$$e_0^\mu = \left(C^{-\frac{1}{2}}, 0, 0, 0\right), e_1^\mu = \left(0, F^{-\frac{1}{2}}, 0, 0\right), e_2^\mu = \left(0, 0, H^{-\frac{1}{2}}, 0\right), e_3^\mu = \left(0, 0, 0, R^{-\frac{1}{2}}\right).$$

(19)

The nonzero components of the covariant derivatives of vierbein vectors can be found using Eqs. (7) and (19):

$$e_{0t;r}, e_{0r;t}, e_{0\theta;t}, e_{1t;t}, e_{1\theta;r}, e_{1\theta;\theta}, e_{1\varphi;\varphi}, e_{2t;t}, e_{2r;r}, e_{2r;\theta}, e_{2\varphi;\varphi}, e_{3r;\varphi}, e_{3\theta;\varphi}.$$

(20)

After using Eqs. (4), (6), (9), (19) and (20) we arrive at the following expressions for the analogues of the electric and magnetic fields:

$$\vec{E} = \left(\frac{\gamma}{\sqrt{F}} e_{0r;t}, \frac{\gamma}{\sqrt{H}} e_{0\theta;t}, 0\right),$$
$$\vec{B} = \left(-e_{2\varphi;\varphi} e_3^\varphi U_\varphi, -e_{3r;\varphi} e_1^r U_\varphi, -[e_{1\theta;r} e_2^\theta U_r + e_{1\theta;\theta} e_2^\theta U_\theta]\right).$$

(21)

The trajectories of particles and photons in the Kerr metric have been studied in [10]. For motion in the equatorial plane, the equations of motion in Eq.(11) are reduced to:

$$\frac{d^2\varphi}{dp^2} + 2\Gamma_{r\varphi}^\varphi \frac{dr}{dp}\frac{d\varphi}{dp} = 0,$$

$$\frac{d^2 t}{dp^2} + 2\Gamma_{tr}^t \frac{dt}{dp}\frac{dr}{dp} = 0,$$

$$\frac{dr^2}{dp^2} + \Gamma_{tt}^r \left(\frac{dt}{dp}\right)^2 + \Gamma_{rr}^r \left(\frac{dr}{dp}\right)^2 + \Gamma_{\varphi\varphi}^r \left(\frac{d\varphi}{dp}\right)^2 = 0.$$

(22)

By solving these equations we find:

$$\gamma^{-1} = \sqrt{\frac{(2M-r)(-2Ma^2+r(6M^2-5Mr+r^2))}{r(a^2M-r(-2M+r)^2)}},$$
$$v_\varphi = \sqrt{\frac{M}{(-Ma^2+r(-2M+r)^2)}}.$$

(23)

After substituting (3) and (21) in Eq. (14) and using Eq. (23), and making change of variables $a = 2yM = \frac{J}{M}, r = 2Mk$, the nonzero component $\Omega_2$ is found to be:

$$\Omega_2 = \frac{\left[\sqrt{\frac{1}{kM-k^2M}}\sqrt{\frac{k^2}{-k+k^2+y^2}}[k[-3\beta-2\zeta]-2k^3[\beta+\zeta]+k^2[5\beta+4\zeta]+y^2[2\beta+\zeta]]\right]}{[8\beta k^4 M \sqrt{\frac{-2k+4k^2-2k^3+y^2}{(-1+k)kM(-k+k^2+y^2)}}[\beta+\zeta]]},$$

(24)

where:

$$\zeta = \sqrt{\frac{(-1+k)(3k-5k^2+2k^3-2y^2)}{k(2k-4k^2+2k^3-y^2)}}, \quad \beta = \sqrt{\frac{k-1}{k}}.$$

In Fig.(3) we show $2|\Omega_2|M$ as a function of $\frac{r}{2M}$ for $0 \leq y \leq 0.5$. We stop at $y = 0.5$ since the Kerr metric does not describe a black hole due to the existence of a naked singularity at $r = 0$.

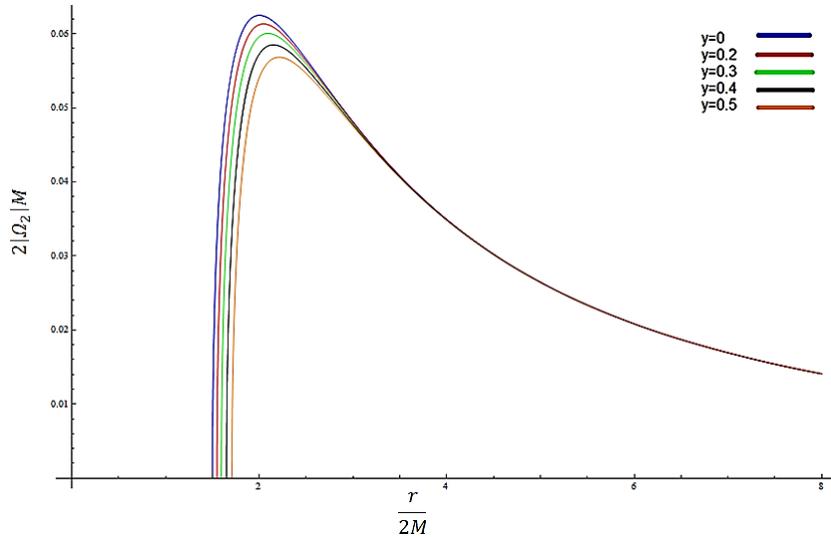

Figure 3: Neutrino spin oscillation frequency versus the radius of the neutrino orbit for different values of $y = \frac{J}{2M^2}$.

It is seen that the maximum frequency decreases with increasing $J$. It is also seen that the frequency has its largest value for $J = 0$. Also, at long distances all the curves approach a common value. This indicates that the effect of rotation diminishes at large $r$. We note that the $y = 0$ case reproduce the result for the Schwarzschild metric [5].

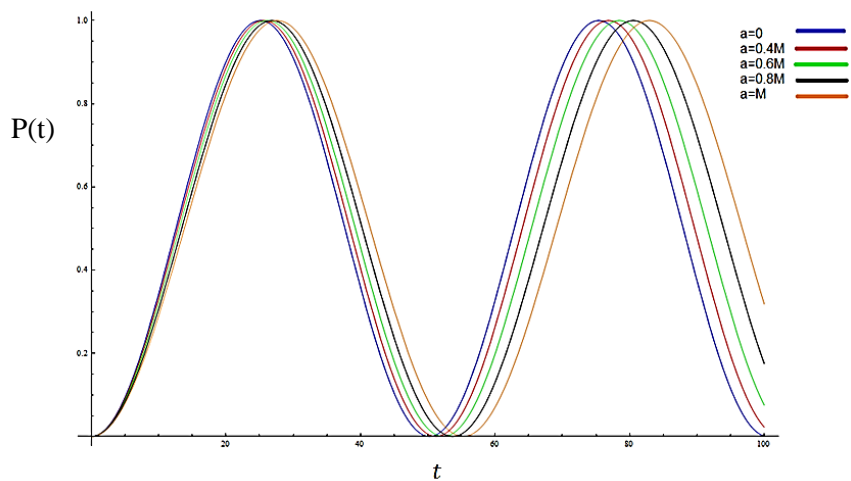

Figure 4: The transition probability versus time for different values of $a$.

In Fig.(4), we show the transition probability $P(t)$.

### Application

We now present a phenomenological application of our results. We consider the simple bipolar neutrino system which is an important example of the collective neutrino oscillations. This system consists of a homogeneous and isotropic gas of initially mono-energetic $\nu_e$ and $\bar{\nu}_e$, discussed in [11,12]. A simple description by the flavor pendulum helps us to understand many qualitative features of the collective neutrino oscillations in supernovae. We introduce $\epsilon$ as the ratio of neutrinos to antineutrinos $\epsilon = \frac{n_\nu}{n_{\bar{\nu}}}$. Consider two such systems, one in the presence of a charged black hole ($Q \leq M$) and the other in the presence of a neutral black hole, with the same initial value of $\epsilon$, $\epsilon_{t=0}^{Q \leq M} = \epsilon_{t=0}^{Q=0}$.

It is seen from Fig. (2) that the period of transition probability is smaller for a charged black hole :

$$T_{Q \leq M} < T_{Q=0}. \tag{25}$$

For Majorana neutrinos, the neutrino spin flip is equivalent to neutrino-antineutrino oscillations. Therefore, at $t > 0$, we have :

$$\epsilon_t^{Q \leq M} < \epsilon_t^{Q=0}. \tag{26}$$

This implies that the precession frequency $\Omega$ of the flavor pendulum as a function of the neutrino number density will be higher for a charged black hole.

Similarly, it is seen from Fig.(4) that:

$$T_{J=0} < T_{J>0}, \tag{27}$$

which results in :

$$\epsilon_t^{J=0} < \epsilon_t^{J>0}. \tag{28}$$

This implies that the precession frequency $\Omega$ of the flavor pendulum as a function of the neutrino number density will be higher for a non-rotating black hole.
One comment is in order at this point. Collective oscillations are known to happen for neutrinos that leave a supernova mainly along radial trajectories. On the other hand, we have studied spin oscillations of neutrinos trapped on circular orbits. To know the reason we have provided below the three components of the frequency for the Reissner-Nordstrom metric :

$$\Omega_1 = \frac{G_1}{\gamma} = \frac{\frac{1}{2}\left[B_1 + \frac{1}{1+u^0}[\vec{E} \times \vec{u}]_1\right]}{\gamma} = \frac{1}{2} v_\varphi \cos\theta,$$

$$\Omega_2 = \frac{G_2}{\gamma} = \frac{\frac{1}{2}\left[B_2 + \frac{1}{1+u^0}[\vec{E} \times \vec{u}]_2\right]}{\gamma} = \frac{v_\varphi}{2} \sin\theta \left(-A + \frac{\gamma}{2(1+\gamma A)}\left(\frac{2M}{r^2} - \frac{2Q^2}{r^3}\right)\right),$$

$$\Omega_3 = \frac{G_3}{\gamma} = \frac{\frac{1}{2}\left[B_3 + \frac{1}{1+u^0}[\vec{E} \times \vec{u}]_3\right]}{\gamma} = \frac{v_\theta}{2}\left(A - \frac{\gamma}{2(1+\gamma A)}\left(\frac{2M}{r^2} - \frac{2Q^2}{r^3}\right)\right).$$

(29)

We see that there will be no spin oscillations in the case of a radial trajectory. It is also pointed out in [5] that there is no spin flip for radial motion in the Schwarzschild metric. On the other hand, in the Kerr metric spin oscillations occur for both radial and circular trajectories. Therefore spin oscillations, which can be considered as neutrino-antineutrino oscillations for Majorana neutrinos is a relevant effect for trapped neutrinos.

## Conclusion

In this paper we have investigated neutrino spin oscillations in the gravitational field of charged and rotating black holes. We have also analyzed the dependence of the frequency of oscillations on the orbit radius. In the case of a charged black hole, the maximum frequency is a monotically increasing function of the charge. For a rotating black hole, the maximum frequency decreases with increasing the angular momentum. We have also briefly studied the effects of charge and rotation of a black hole on a bipolar neutrino system.

## Acknowledgments

We would like to thank Rouzbeh Allahverdi (university of New Mexico) for proof reading of the manuscript.